\documentclass[journal]{IEEEtran}

\usepackage[pdftex]{graphicx}
\graphicspath{{figures/}}
\DeclareGraphicsExtensions{.jpg,.png}

\usepackage[caption=false,font=footnotesize]{subfig}

\usepackage{amsmath}
\usepackage{bbm}
\usepackage{stmaryrd}

\usepackage{array}

\usepackage[hyphens]{url}
\usepackage[pdftitle=Title,pdfauthor=Anonymous]{hyperref}
\usepackage{adjustbox}
\usepackage{booktabs}
\usepackage{multirow}
\usepackage{rotating}
\usepackage{makecell}
\usepackage{hhline}
\usepackage{lscape}
\usepackage{tabu}
\usepackage{tikz}
\usepackage{textcomp} 
\usetikzlibrary{shapes,backgrounds} 
\usepackage[justification=centering]{caption} 
\usepackage{colortbl}

\usepackage{blindtext}
\usepackage[utf8]{inputenc}
\usepackage[T1]{fontenc}
\usepackage{amsmath}
\usepackage{amsfonts}
\usepackage{amssymb}
\usepackage{tabularx}
\usepackage{blindtext}
\usepackage{booktabs}
\usepackage{makecell}

\usepackage{cellspace}

\usepackage{longtable}

\usepackage{listings, xcolor}

\definecolor{verylightgray}{rgb}{.97,.97,.97}
\lstdefinelanguage{Solidity}{
keywords=[1]{anonymous, assembly, assert, balance, break, call, callcode, case, catch, class, constant, continue, contract, debugger, default, delegatecall, delete, do, else, event, export, external, false, finally, for, function, gas, if, implements, import, in, indexed, instanceof, interface, internal, is, length, library, log0, log1, log2, log3, log4, memory, modifier, new, payable, pragma, private, protected, public, pure, push, require, return, returns, revert, selfdestruct, send, storage, struct, suicide, super, switch, then, this, throw, transfer, true, try, typeof, using, value, view, while, with, addmod, ecrecover, keccak256, mulmod, ripemd160, sha256, sha3}, 
	keywordstyle=[1]\color{blue}\bfseries,
	keywords=[2]{Stages,States, address, bool, byte, bytes, bytes1, bytes2, bytes3, bytes4, bytes5, bytes6, bytes7, bytes8, bytes9, bytes10, bytes11, bytes12, bytes13, bytes14, bytes15, bytes16, bytes17, bytes18, bytes19, bytes20, bytes21, bytes22, bytes23, bytes24, bytes25, bytes26, bytes27, bytes28, bytes29, bytes30, bytes31, bytes32, enum, int, int8, int16, int24, int32, int40, int48, int56, int64, int72, int80, int88, int96, int104, int112, int120, int128, int136, int144, int152, int160, int168, int176, int184, int192, int200, int208, int216, int224, int232, int240, int248, int256, mapping, string, uint, uint8, uint16, uint24, uint32, uint40, uint48, uint56, uint64, uint72, uint80, uint88, uint96, uint104, uint112, uint120, uint128, uint136, uint144, uint152, uint160, uint168, uint176, uint184, uint192, uint200, uint208, uint216, uint224, uint232, uint240, uint248, uint256, var, void, ether, finney, szabo, wei, days, hours, minutes, seconds, weeks, years},	
	keywordstyle=[2]\color{teal}\bfseries,
	keywords=[3]{block, blockhash, coinbase, difficulty, gaslimit, number, timestamp, msg, data, gas, sender, sig, value, now, tx, gasprice, origin},	
	keywordstyle=[3]\color{violet}\bfseries,
	identifierstyle=\color{black},
	sensitive=false,
	comment=[l]{//},
	morecomment=[s]{/*}{*/},
	commentstyle=\color{gray}\ttfamily,
	stringstyle=\color{red}\ttfamily,
	morestring=[b]',
	morestring=[b]"
}

\begin{document}
%
\title{Recent Advancements in Mode Division Multiplexing for Communication and Computation in Silicon Photonics}
%
%
%

\author{Kaveh~Rahbardar~Mojaver,~\IEEEmembership{Member,~IEEE,}
        Seyed~Mohammad~Reza~Safaee,~\IEEEmembership{Student~Member,~IEEE,}
        Sunami~Sajjanam~Morrison,~\IEEEmembership{Student~Member,~IEEE,}
        and~Odile~Liboiron-Ladouceur,~\IEEEmembership{Senior~Member,~IEEE}
\thanks{All the authors are with the Department
of Electrical and Computer Engineering, McGill University, Quebec,
Canada,  H3A 0E9, e-mail: mojaver@ieee.org, seyed.safaeeardestani@mail.mcgill.ca, sunami.sajjanammorrison@mail.mcgill.ca, and odile@ieee.org}
\thanks{Manuscript received 1 March 2024.}}

%
%

\markboth{Journal of Lightwave Technology,~Vol.~xx, No.~yy, March~2024}%
{Shell \MakeLowercase{\textit{et al.}}: Bare Demo of IEEEtran.cls for IEEE Journals}
%



\maketitle

\begin{abstract}
Mode Division Multiplexing (MDM) is a technique used over the past decade in Silicon Photonics (SiPh) to incorporate more data into communication links by employing higher-order transverse electric or transverse magnetic modes. MDM was primarily used in optical communication; however, in recent years, there have been several applications of MDM in optical computing, including both classical and quantum computing. Although MDM has shown great promise for increasing the throughput of optical communication and the accuracy and fidelity of optical computation, there are a few challenges towards expanding its applications. One major challenge is the lack of process design kits (PDKs) and building block libraries compatible with standard SiPh foundries. Here, we present a comprehensive library of MDM components developed using classical and inverse design, compatible with standard 220 nm SiPh foundries. The library includes thermo-optic phase shifters, mode multiplexers and demultiplexers, mode converters, mode exchangers, and multi-mode interference couplers. We also discuss our recent achievements in MDM for datacom, classical and quantum optical computing, including a mode-selective switch for mode-reconfigurable optical add-drop multiplexer (ROADM), multimode multiply-accumulate operation, and multimode photonic quantum processors.
\end{abstract}


%
\IEEEpeerreviewmaketitle

\section{Introduction}
%
%
%
%
\IEEEPARstart{M}{ode} Division Multiplexing (MDM) is a technique that incorporates multiple spatial optical modes inside an optical waveguide or fiber to increase the capacity of communication channels. Introduced over a decade ago in Silicon Photonics (SiPh), MDM was primarily aimed at enhancing the throughput of on-chip communications \cite{Dai:Mux, MDM_2014_Lipson} but has gradually expanded to chip-to-chip communication through fibers as well \cite{MDM_fiber_11,MDM_fiber_11_2,Lipson_MDM_fiber,Vuckovic_mdm}. Transverse electric (TE) and transverse magnetic (TM) modes are solutions to the Maxwell's equations for the energy distribution of electromagnetic waves propagating inside a waveguide or fiber. These modes are orthogonal; hence, they do not interfere with each other and can be used as separate communication channels.

The key advantage of MDM over wavelength division multiplexing (WDM) lies in its ability to use a single laser for all channels, significantly saving power, as lasers constitute the most power-hungry component in communication links \cite{MDM_laser_power}. Additionally, in MDM, mode conversion can be easily achieved through energy redistribution inside a waveguide \cite{Masnad:Mode_Converter, Zhang:mode_exchanger}, while WDM links require nonlinear effects to alter the wavelength of photons. Importantly, MDM should not be viewed as a competitor to WDM but rather as a complement, offering increased degrees of freedom for scaling optical systems \cite{Rubana_JLT, Rubana_JSTQE}.

Despite more than a decade since its inception, several challenges hinder the widespread adoption of MDM. The primary and most significant challenge is modal crosstalk, where any factor affecting energy distribution inside a waveguide, such as sidewall roughness or fabrication variations, can lead to modal crosstalk \cite{Chris:crosstalk}. Efficient coupling in and out of the chip also remains challenging in MDM \cite{Dai_Mode_converter_coupler}. Additionally, since higher-order spatial modes are less confined in the waveguide, they require larger radius bends, resulting in a larger footprint \cite{Dai_bends}. Modal crosstalk is exacerbated by waveguide crossings, making complex designs with multiple crossings challenging in MDM \cite{MDM_Crossing}. Another challenge is that the process design kits (PDK) of most standard SiPh foundries include only single-mode components, necessitating the development of the required building blocks for MDM system design—an especially daunting task for research-level and small-scale industry designers. Even most commercial system-level optical simulators do not support MDM, as standard PDKs lack MDM components.

Despite these challenges, recent advancements in MDM have showcased its potential in on-chip and chip-to-chip communication. The coupling of three TE and one TM modes from a SiPh multimode waveguide to a few-mode fiber achieved an average crosstalk of -7 dBm, enabling data transmission at a rate of 1.92 Tb/s in the L-band via MDM \cite{Lipson_MDM_fiber}. Furthermore, a successful demonstration of 1.12 Tb/s data transmission via a multimode rectangular-shaped fiber using MDM and WDM was achieved with inverse design fiber-to-chip couplers \cite{Vuckovic_mdm}. Beyond multiplexing data, it has been shown that higher-order modes can be used to carry the clock or local oscillators in source synchronous communication systems \cite{Chris_williams}. Based on a similar approach, a 7.2-Tb/s self-homodyne coherent transmission over a weakly coupled few-mode fiber using 10 spatial modes has recently been demonstrated, where nine modes were used to carry the data, and one mode is used to carry the local oscillator \cite{10mode_MDM}.

Beyond optical communication, MDM has garnered interest in optical computation. Higher-order modes prove useful in enhancing programming speed and accuracy in programmable optical processors for energy-efficient vector matrix multiplication widely used in machine learning (ML) and artificial intelligence (AI) \cite{Kaveh_JSTQE}. The orthogonal modes allow for non-coherent summation of optical energy, finding applications in multimode Multiply-Accumulate (MAC) operations to increase the number of bits and overcome coherent summation challenges. Moreover, different spatial modes can be employed to encode quantum states of light in quantum photonics, facilitating the scalability of optical quantum systems \cite{DAI_QUANTUM}.

In this paper, we start with a brief review of the fundamentals of MDM in SiPh. Next, we provide a library of MDM building blocks compatible with standard 220 nm SiPh that can be used by photonic designers to develop various MDM designs. Then we discuss recent advancements in MDM in datacom, and optical classical and quantum computing.

\section{MDM: Fundamentals}

\begin{figure}[!t]
\centering
\includegraphics[width=8.5cm]{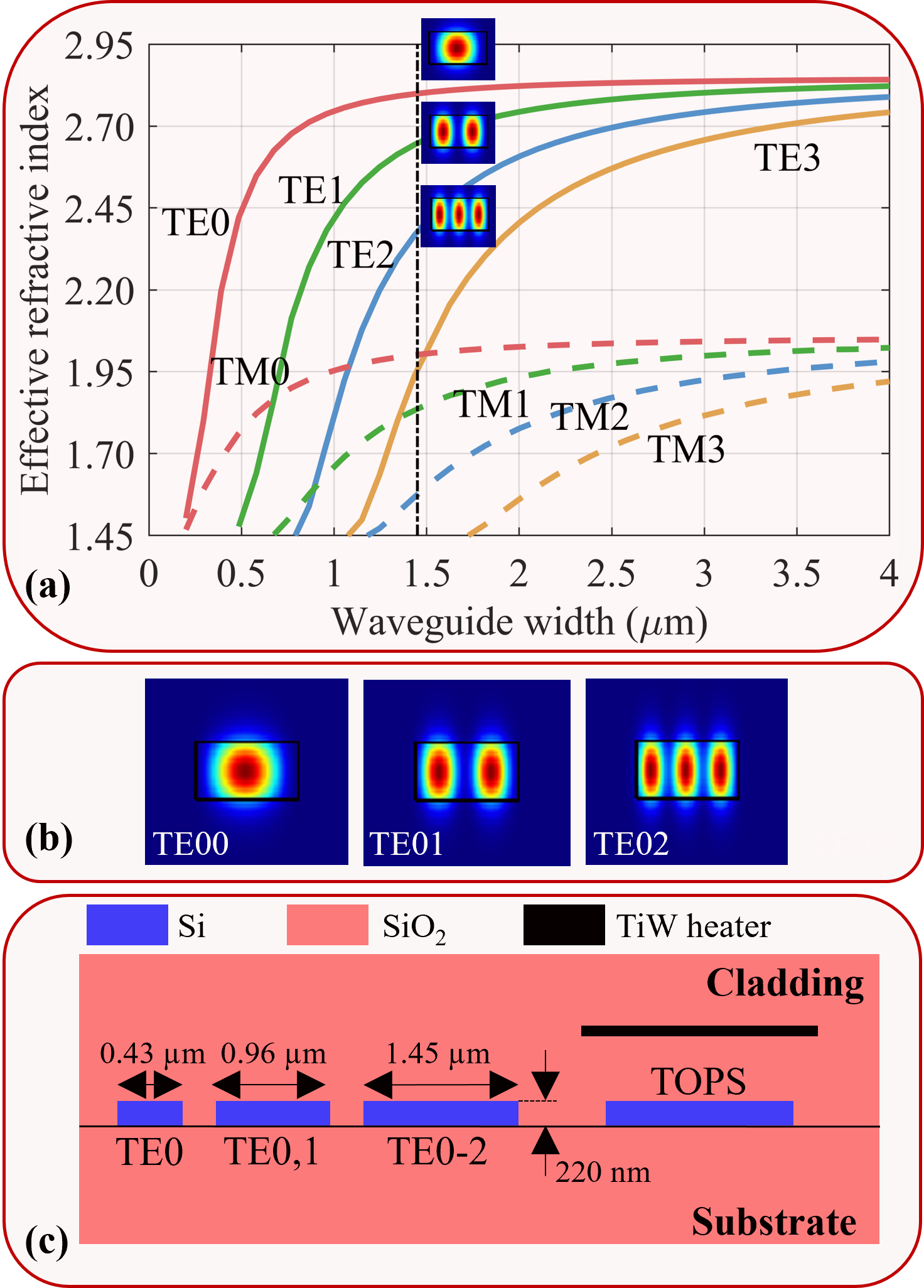}
\caption{(a) Calculated effective refractive index of a 220 nm thick strip SiPh waveguide versus its width for the first four TE and TM modes. (b) Profiles (energy density) of the first three TE modes for a 1.45 \textmu m width case. (c) Cross section view of 220~nm SiPh process.}
\label{fig_neff}
\end{figure}

Figure \ref{fig_neff} (a) displays the computed effective refractive index, $n_{\text{eff}}$, of the first four TE and TM modes in a 220 nm thick strip waveguide at a wavelength of 1550 nm, plotted against the waveguide width. For all modes, $n_{\text{eff}}$ increases with the waveguide width. Modes with $n_{\text{eff}}$ surpassing the refractive index of the SiO\textsubscript{2} cladding remain confined and guided within the waveguide. For instance, with a 430 nm waveguide width, only the fundamental TE mode is guided, classifying it as a single mode. However, the number of TE modes increases with the waveguide width. For a 1.45 \textmu m waveguide, the first three TE modes (TE0, TE1, and TE2) are guided. Figure~\ref{fig_neff}~(b) displays the cross section of a 1.45 \textmu m by 220 nm SiPh waveguide and the mode profile of the first three TE modes inside this waveguide. Figure \ref{fig_neff}~(c) shows a cross-section of a typical 220 nm thick SiPh process, including single-mode and multimode waveguides, as well as thermo-optic phase shifters (TOPS).

The TE and TM modes, serving as orthogonal solutions to the Maxwell equations within the waveguide, manifest as separate channels suitable for multiplexing different signals in an MDM system. Optical modes depict the distribution of electromagnetic wave energy within the waveguide, enabling a relatively straightforward conversion between them, a double-edged sword. On one hand, the advantage lies in the ease of converting between two MDM channels by redistributing the energy within the waveguide, without relying on nonlinearity effects as required for WDM technology. Section III will showcase mode converters designed using an inverse design approach. On the other hand, however, a significant drawback arises in the presence of waveguide sidewall roughness, where energy redistribution can lead to unwanted coupling between modes, resulting in modal crosstalk. This issue stands as a key challenge to be addressed in MDM systems.

\captionsetup{%
    justification=justified,%
}

\section{MDM building blocks}
The PDK from most standard SiPh foundries, such as Interuniversity Microelectronics Centre (IMEC), Advanced Micro Foundry (AMF), and Global Foundries (GF), typically includes only single-mode components. Consequently, designing multimode photonic systems poses a challenge, as designers often need to create all components, including fundamental ones like splitters and combiners. To address this issue, we present an open-access SiPh library dedicated to multimode components \cite{PDK, Kaveh_GLSVLSI}. This comprehensive library comprises essential building blocks such as Multimode Interferometers (MMIs), mode (de)multiplexers, mode-sensitive TOPS, and mode-insensitive TOPS, catering to both classical and quantum optical computing in multimode systems. Notably, this library seamlessly integrates with the standard 220 nm thick SiPh technology platform available through commercial microfabrication foundries.
\subsection{MDM Multimode Interferometers}

MMIs are widely employed in integrated optics for splitting and combining optical signals. In an MDM system, an MMI capable of simultaneously splitting/combining all the desired modes is necessary.

\begin{figure*}[!t]
\centering
\includegraphics[width=17cm]{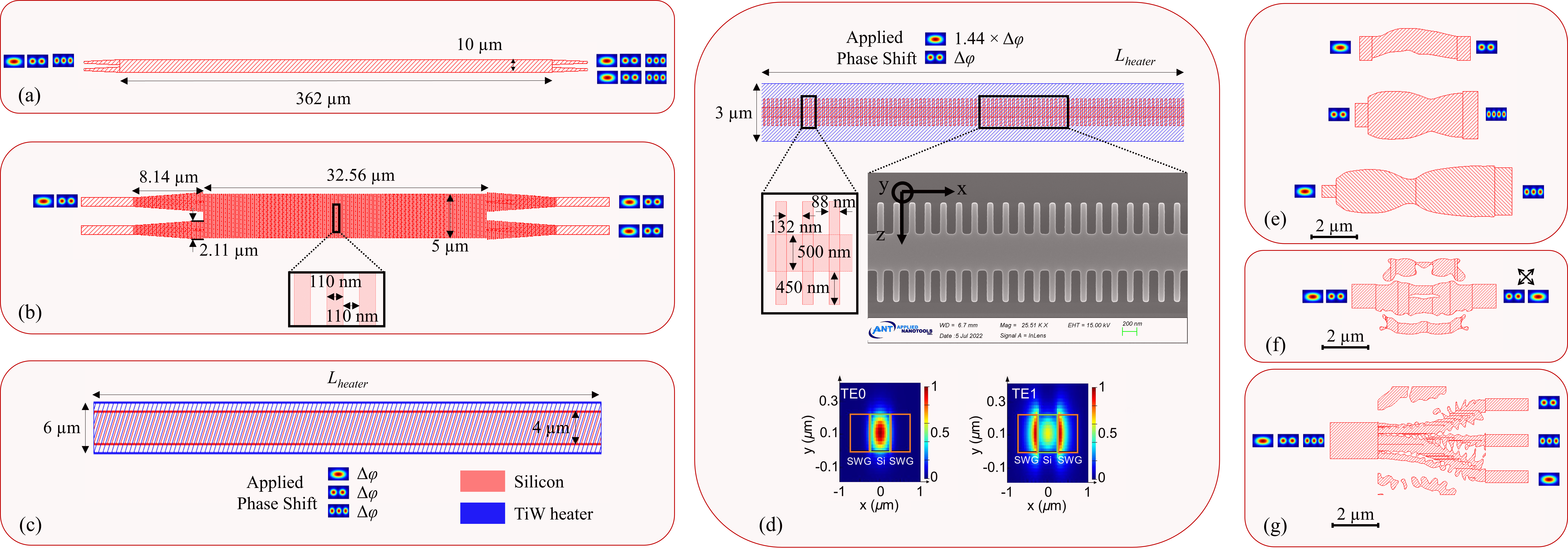}
\caption{
Library of MDM building blocks available through the online library \cite{PDK}: (a) MMI that splits/combines the first three TE modes (TE0, TE1, TE2) using conventional design \cite{Zhang:20, Alok_MDM}. (b) SWG-based MMI with a compact footprint for the first two TE modes \cite{Hatef_MMI}. (c) Mode-insensitive thermo-optic phase shifter (MI-TOPS) that applies the same phase shift to the first three TE modes \cite{Zhang:20}. (d) SWG-based mode-selective thermo-optic phase shifter (MS-TOPS) that applies a different phase shift to TE0 and TE1. The mode profile shows TE0 propagating in the core of the phase shifter made by Silicon, while TE1 has two side lobes propagating in the SWG region with different temperature coefficients \cite{JLT_Mohammad}. (e) Compact mode converters enabled using inverse design for converting between various TE modes \cite{Masnad:Mode_Converter}. (f) Inverse-designed mode exchanger for swapping between TE0 and TE1 \cite{Zhang:mode_exchanger}. (g) Inverse-designed mode (de)Mux for the first three TE modes. The design is optimized to compensate for fabrication variations using PreFab \cite{Dusan_ECOC22, Dusan_prefab}.
}
\label{fig_PDK}
\end{figure*}
Figure \ref{fig_PDK}~(a) illustrates a mode-insensitive MMI designed for the first three TE modes (TE0, TE1, TE2). This MMI performs a 50:50 split/combine of all these modes simultaneously. To achieve the desired two-fold image of the three input modes, the multimode MMI width is selected to be 10 µm to support a sufficient number of modes in the multimode region. The beat length ($L_\pi$) is initially calculated using:
\begin{eqnarray}
L_\pi = \frac{\pi}{{\beta_0 - \beta_1}} = \frac{{4n_{\text{eff}} W_m^2}}{{3\lambda_0}}
\end{eqnarray}
where, $\beta_0$ and $\beta_1$ represent the propagation constants of the first two TE modes within the MMI. The effective index, denoted as $n_{\text{eff}}$, the width of the MMI, labeled as $W_m$, and the desired wavelength is $\lambda_0$. To optimize the MMI dimensions, we leverage commercial numerical simulation-based software like Ansys/Lumerical. Following this optimization process, the determined length of the MMI is 362 \textmu m, tailored for efficient operation at a wavelength of 1550 nm. Details of the design are presented in \cite{Zhang:20, Alok_MDM}. 
While highly efficient, the conventional MMI exhibits a substantial footprint. Figure \ref{fig_PDK}~(b) illustrates the layout of a broadband compact dual-mode MMI utilizing subwavelength grating (SWG) structures. The SWG-based MMI is nearly ten times more compact than its conventional counterpart, showcasing a flat response with low loss and crosstalk over a bandwidth of 100~nm. Excess loss measurements reveal values below 0.1 dB and 0.65 dB for TE0 and TE1, respectively, with modal crosstalk measuring less than -17 dB for both input modes. Comprehensive design details and experimental validation for this SWG-based MMI can be found in \cite{Hatef_MMI}.

In the microfabrication process of SWG structures, the SiO$_2$ cladding deposition may not completely fill the gaps between the fine features of the SWG, resulting in residual air gaps. To further enhance the MMI's specifications, particularly in addressing modal crosstalk, and to mitigate the impact of fabrication variations, an investigation into the influence of buried air gaps in the cladding of SiPh structures is conducted and the design is optimized for the best performance in the presence of air gaps \cite{Hatef_voids}.

\subsection{MDM Thermo-Optic Phase Shifters}
Applying phase shift to the light is essential in several applications including switching, sensing, and computation. Employing TOPS's is the most common method for applying phase shift to optical waves in integrated optics. A TOPS typically comprises a waveguide positioned underneath a metal heater. The heater applies specific heat corresponding to its bias voltage (or bias current) onto its underneath waveguide. The refractive index of the waveguide varies with its temperature, and this alteration in the refractive index results in a change in the phase of light as it travels through the waveguide. The temperature-dependent phase change in the optical waveguide is \cite{Jacques:19}:
\begin{eqnarray}
\Delta \phi = \frac{2 \pi L}{\lambda_0} \frac{dn}{dT} \Delta T
\end{eqnarray}
where $L$ is the TOPS length, $\lambda_0$ is the free-space wavelength of light, $\Delta T$ is the temperature change, and \textit{dn/dT} is the thermo-optic coefficient which is approximately $1.86 \times 10^{-4} K^{-1}$ for Silicon \cite{temp_coeff}.
In MDM systems different modes have different propagation constants and effective refractive indices, hence, for an specific TE\textsubscript{x} (or TM\textsubscript{x}) modes with an effective refractive index of $n_\text{eff}$, we rewrite equation eq. 2 as:
\begin{eqnarray}
\Delta \phi_{TE_x} = \frac{2 \pi L}{\lambda_0} \frac{dn_{eff}}{dT} \Delta T.
\end{eqnarray}

Simulation results using a commercial CAD tool from Ansys Lumerical shows the change in the rate of the effective index with respect to the temperature ($dn_{\text{eff}}/dT$) converges toward almost the same value for different TE modes as the TOPS width increases \cite{Zhang:20}. According to this simulation for a TOPS with the waveguide width higher that 4 \textmu m the difference between the $dn_{\text{eff}}/dT$ for the first three TE modes is smaller than $2\%$. Therefore, a TOPS with the waveguide width higher that 4 \textmu m is considered \textit{mode-insensitive} for the first three TE modes (Figure \ref{fig_PDK}~(c)). The application of mode-insensitive TOPS (MI-TOPS) is mainly where the same phase shift to all the modes are desired. For example, in a multimode communication system, when different modes carry different multiplexed signals, if routing all multiplexed channels in a same direction is desired, space switches with mode insensitive phase shifters are used \cite{Zhang:20}.

In contrast, several applications including multi-transverse-mode quantum gates, multi-transverse-mode optical processor (MTMOP) \cite{Kaveh_JSTQE}, and mode selective optical switches need mode-selective TOPS (MS-TOPS) offering widely different thermo-optic coefficients for different modes. Specifically, where routing a specific mode to a desired path is needed, a MS-TOPS is beneficial. We showed mode-selective operation of a narrow  TOPS with a width smaller than 1 µm \cite{Zhang:20}, \cite{Kaveh_JSTQE}. However, the mode selectivity of this TOPS is relatively poor resulting in less than 1.1 times difference in the thermo-optic coefficient between TE0 and for TE1 (1.8 and 1.96 for TE0 and TE1, respectively.) 

We have developed a MS-TOPS with a distinct separation between the thermo-optic coefficients of TE0 and TE1, achieved through the incorporation of SWG structures \cite{JLT_Mohammad}. Figure~\ref{fig_PDK}~(d) displays the scanning electron microscope (SEM) image of the SWG-based MS-TOPS, providing insight into its design parameters. The proposed MS-TOPS design relies on the fact that TE0 primarily propagates within the 500 nm waveguide center, while the TE1 field pattern exhibits two lobes closer to the waveguide sidewalls. Silicon's thermo-optic coefficient is approximately an order of magnitude larger than that of oxide, specifically $1.86 \times 10^{-4} K^{-1}$ for silicon and $0.95 \times 10^{-5} K^{-1}$ for SiO\textsubscript{2} \cite{temp_coeff}. By engineering the thermo-optic coefficient of the regions near the waveguide sidewalls, while maintaining the waveguide center as silicon, a substantial difference in thermo-optic coefficients between TE0 and TE1 can be achieved.

The SWG structures feature a duty ratio defined as the length of the Silicon over the pitch length. For the MS-TOPS shown in fig.~\ref{fig_PDK}~(d), the duty ratio is 40\% since the width of Silicon in the periodic structure is 88~nm and the pitch is 220~nm. The Finite-Difference Eigenmode (FDE) Lumerical mode solver is employed to calculate the effective refractive index of the MS-TOPS for TE0 and TE1 \cite{IPC_22_MSTOPS, SiPh_23_MSTOPS}. The SWG section of the MS-TOPS is modeled as a homogeneous material with an equivalent refractive index. Simulation results presented in fig.~\ref{fig_PDK}~(d) illustrate the TE0 main lobe predominantly in the silicon core and the two lobes of TE1 in the SWG side structures. Experimental results demonstrate a mode-selectivity of 1.44 for a duty ratio of 40\%, signifying that the thermo-optic coefficient of TE0 is 44\% larger than that of TE1 \cite{JLT_Mohammad}. A cascaded configuration of the proposed MS-TOPS and a MI-TOPS provides two degrees of freedom to manipulate the relative phase of each mode independently as it can be seen in the next sections in developing mode selective switches.

\subsection{Mode converters and mode exchangers}
Mode converters and mode exchangers are essential in MDM communications when changing or swapping channels is required. In multi-mode computations, they can be used to convert optical data encoded on a specific transverse mode to another mode. Unlike WDM, which requires nonlinearity to convert the wavelength, MDM allows for simple designs with small footprints to convert or exchange transverse modes. Computational inverse design techniques have shown potential for efficiently designing mode converters and exchangers with minimal footprints. TE mode converters on the silicon-on-insulator platform have been experimentally demonstrated using the computationally efficient shape optimization method \cite{Masnad:Mode_Converter}. These mode converters, shown in fig.~\ref{fig_PDK}~(e), exhibit conversion efficiencies above 95\%, with an insertion loss ranging from 0.3 dB to 1 dB over a wavelength of 1.5 µm to 1.58 µm. They have been experimentally validated in the time domain with a 28 Gbps non-return-to-zero (NRZ) and 20 GBaud  4-level pulse-amplitude modulation (PAM-4) payload transmissions and have been added to the MDM library.

Mode exchangers can exchange between multiple modes; for example, they can swap TE0 and TE1 or even higher-order modes. A compact 4 \textmu m $\times$ 4 \textmu m footprint mode exchanger that can swap TE0 and TE1 is designed using inverse design and shown in Fig. \ref{fig_PDK}~(f). This mode exchanger is demonstrating 0.3 dB insertion loss and -29.5 dB crosstalk at 1550 nm \cite{Zhang:mode_exchanger}. Mode exchangers for higher-order modes, which exchange TE0 mode with TE1 mode as well as TE2 mode with TE3 mode, are also presented in \cite{Zhang:mode_exchanger}. As we will discuss in the next sections, these mode exchangers can be used in Transverse-Mode Encoded Quantum Processors where it is desired to transfer quantum information encoded on one mode to another. They can also be used in optical switches to eliminate the mode dependency of the phase shifters by cascading phase shifters and switching the modes in between \cite{Zhang:mode_exchanger}.

\subsection{Mode (de)Mux}

Conventional mode (de)Mux can be effectively designed by employing adiabatic-based asymmetrical directional couplers for TE and TM modes, as outlined in \cite{Dai:Mux}. In a specific example, an adiabatic directional coupler-based mode multiplexer, featuring insertion losses varying between -0.3 dB to -2.9 dB for the TE0 mode, -0.9 dB to -3.7 dB for the TE1 mode, and -0.3 dB to -2.2 dB for the TE2 modes within the 1520 nm to 1600 nm wavelength range, is demonstrated in \cite{Priti:19} and included in the MDM library \cite{PDK}. Notably, the worst crosstalks for this (de)Mux arrangement are reported as $-18.3$~dB, $-17.0$~dB, and $-19.6$~dB for TE0, TE1, and TE2 modes, respectively.

While adiabatic-based (de)Mux configurations are known for their efficiency, they often come with relatively large footprints. To address this, inverse design methodologies have been employed to create mode (de)Muxs with smaller footprints. For instance, a topologically optimized (de)Mux designed for TE0, TE1, and TE2 modes showcases a compact 4.5~\textmu m~$\times$~4.5~\textmu m footprint as shown in fig. \ref{fig_PDK}~(g). This (de)Mux has 0.13 dB insertion loss and worst-case channel crosstalk is below -18.5 dB across a wavelength bandwidth of 1500 nm-1600 nm \cite{Masnad_IPC}. A significant challenge associated with inverse-designed devices lies in their small footprint, including numerous features with small dimensions. Fabrication variations and imperfections can often lead to performance degradation, particularly in terms of modal crosstalk for MDM components. One viable solution involves employing a deep learning model for predicting fabrication variations in planar integrated silicon photonic devices. A convolutional neural network (CNN) has been demonstrated to learn the relationship between a design and its fabricated outcome, allowing predictions for new, unseen photonic devices based on a small collection of SEM images \cite{Dusan_prefab}. This prediction can then be re-simulated to estimate the expected variation in the experimental performance of a given device. By utilizing this technique and incorporating PreFab fabrication prediction tools \cite{Prefab}, a 3-channel (de)Mux with enhanced performance has been experimentally demonstrated \cite{Dusan_ECOC22}.

\section{MDM in DataCom}

MDM technology has found application in intra-chip communication to enhance data transmission throughput \cite{Zhang:20}. Additionally, it has proven effective in chip-to-chip transmission through the utilization of multi-mode or a few-mode fibers. A recent demonstration showcased the coupling of three TE and one TM modes from a SiPh multimode waveguide to a few-mode fiber, achieving an average crosstalk of -7 dBm. This setup enabled data transmission at a rate of 1.92 Tb/s in the L-band via MDM, complemented by WDM \cite{Lipson_MDM_fiber}. Given the mode mismatch between waveguide and few-mode fibers, a 3D inverse-tapered polymer mode converter was employed at the coupling interface to facilitate mode conversion.

Furthermore, an inverse-designed chip-to-fiber vertical coupler has been developed for the first four TE modes. This coupler facilitates the coupling of modes from a multimode waveguide in SiPh to a multimode fiber with a rectangular core. The shape and energy distribution of the modes within the waveguide and the rectangular core fiber remain largely unchanged, while the mode size is adjusted and optimized to enhance coupling efficiency. Using this technique, a 1.12~Tb/s data transmission via a multimode fiber was successfully demonstrated based on MDM and WDM \cite{Vuckovic_mdm}. This section highlights our recent progress in MDM for datacom.

\subsection{Multimode spatial switches}

\begin{figure}[!t]
\centering
\includegraphics[width=8.5cm]{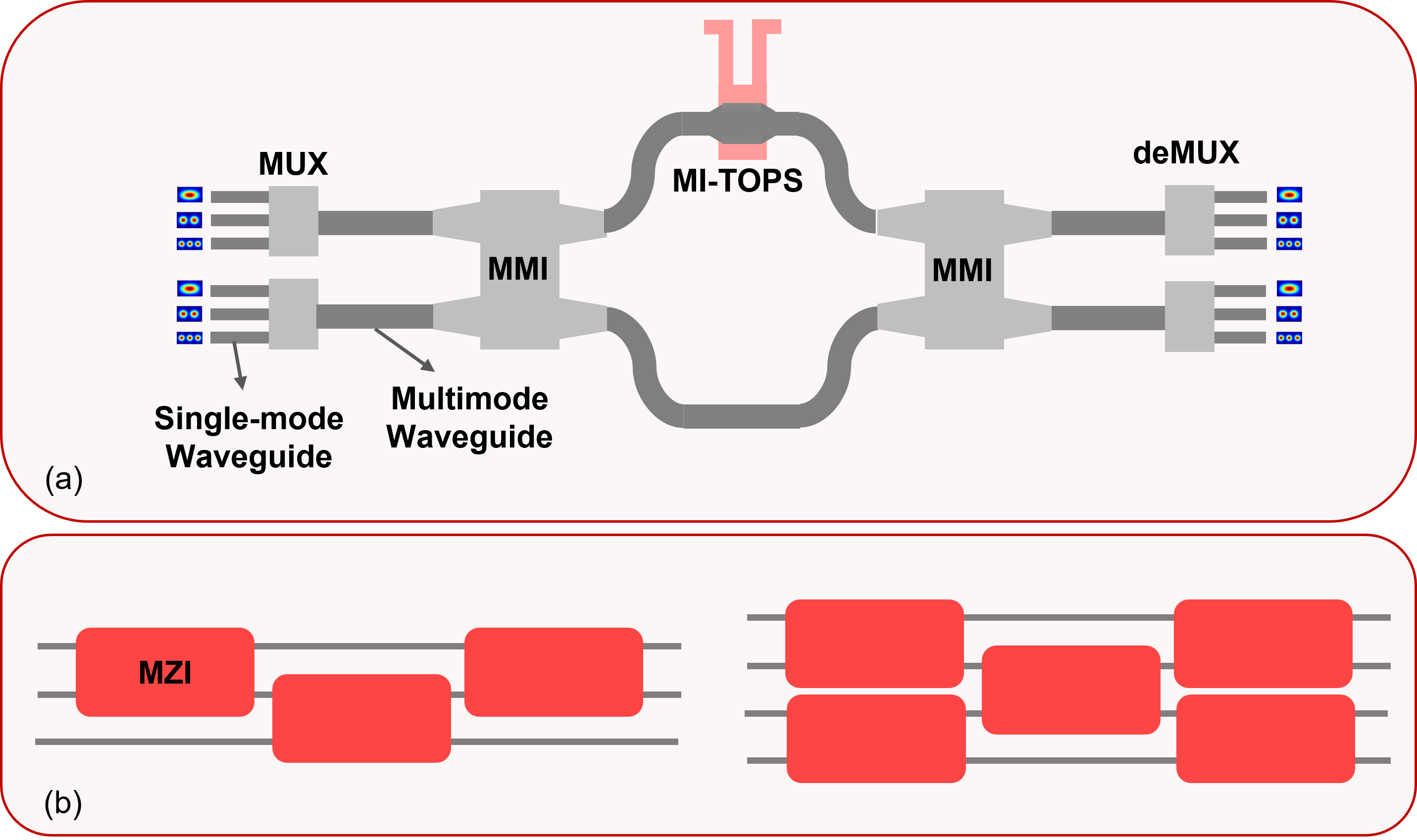}
\caption{(a) Schematic view of a 2 × 2 multimode space switch, including Mux and deMux, and an MZI consisting of two MMIs and one MI-TOPS. (b) Scaling the switch to 3 × 3 and 4 × 4 using Spanke-Bene\u{s} topology.}
\label{fig_MDM_switch}
\end{figure}

Multimode spatial switch is similar to the conventional spatial switch with this difference that each input/output channel carries multiple multiplexed modes instead of a single mode. In fact, the switch can route from different inputs to different outputs while all the multiplexed modes go in the same direction. In other words, the switch features routing multimode signal but does not have the option to selectievly route speceifc modes. Figure \ref{fig_MDM_switch} (a) shows the schematic of a 2 $\times$ 2 MZI-based multimode space switch. The switch includes balanced MMIs (or multimode directional couplers) as multimode splitters and combiners. It also includes a MI-TOPS to apply simmilar phase shift to all the modes, hence, all the multiplexed modes will be routed to the same direction. All the componenets used in this switch are already available in the MDM library \cite{PDK}. This switch can be scaled to higher number of ports employing conventional optical space switch architectures such as Banyan, Bene\u{s}, Spanke-Bene\u{s}, etc. \cite{switch_architecture}.

A three-mode MDM space switch is experimental demonstrated for the TE0, TE1, and TE2 showing approximately -2, -3.7, and -5.2 dB insertion loss for the TE0, E1, and TE2 modes at 1550 nm, respectively. The corresponding crosstalk is less than -8.6 (-9), -8(-10.3), and -10 dB (-10.3 dB) within the wavelength range of 40 nm (1535–1575 nm) for the cross (bar) states, respectively. Error free 10 Gbps PRBS-31 payload transmission over this switch has been demosntrated \cite{Zhang:20}. Using the similar building block, 3 $\times$ 3 and 4 $\times$ 4 MDM switches utilizing TE0 and TE1, shown in fig. \ref{fig_MDM_switch} (b), were experimentally demonstrated in \cite{Alok_MDM}. As multimode waveguide crossing often exhibit large insertion loss and intermodal crosstalk \cite{MDM_Crossing}, Spanke-Bene\u{s} topology is used in the design of this switch to avoid the use of multimode waveguide crossings. The 3-port switch exhibits approximately 2.6 dB and 3.3 dB insertion loss for the longest path with a crosstalk of at most 10 dB and 8 dB over a bandwidth of 40 nm (1530 nm to 1570 nm) for TE0 and TE1 modes, respectively. The insertion loss measured for the 4-port switch is approximately 2.7 dB and 3.6 dB at 1550 nm with a corresponding crosstalk less than 8 dB for the two TE modes, respectively. The suscessful payload transmission using 10 Gb/s NRZ and 14.0625 Gbaud PAM-4 pseudorandom binary sequence (PRBS)-31 data signal has been demosntrated on these switches \cite{Alok_MDM}. A broadband two-mode switch with 60 nm wavelength range from 1530 to 1590 nm has been also demonstrated by optimizing the MMI and phase shifter design over a wider range of wavelength \cite{Guowu_broadband}.

\subsection{Reconfigurable Mode Selective Switch}

A mode-selective switch selectively routes a specific mode without undesirable effects on the others, directing different modes to distinct switch outputs. This switch also enables selective addition or removal of desired channels in a multimode waveguide. Despite considerable advancements in MDM components within SiPh, selectively accessing specific modes without introducing unwanted interference to the others has remained challenging. This challenge becomes more pronounced when attempting to extract the fundamental mode from a multimode bus waveguide while keeping higher-order modes unchanged. Recent approaches have addressed this issue by using modal field redistribution, effectively confining modes to specific regions of the bus waveguide \cite{Dai_ROADM,Dai_mode_selective}. Although efficient, this technique involves redistributing energy among transverse modes within the waveguide. 

\begin{figure}[!t]
\centering
\includegraphics[width=8.5cm]{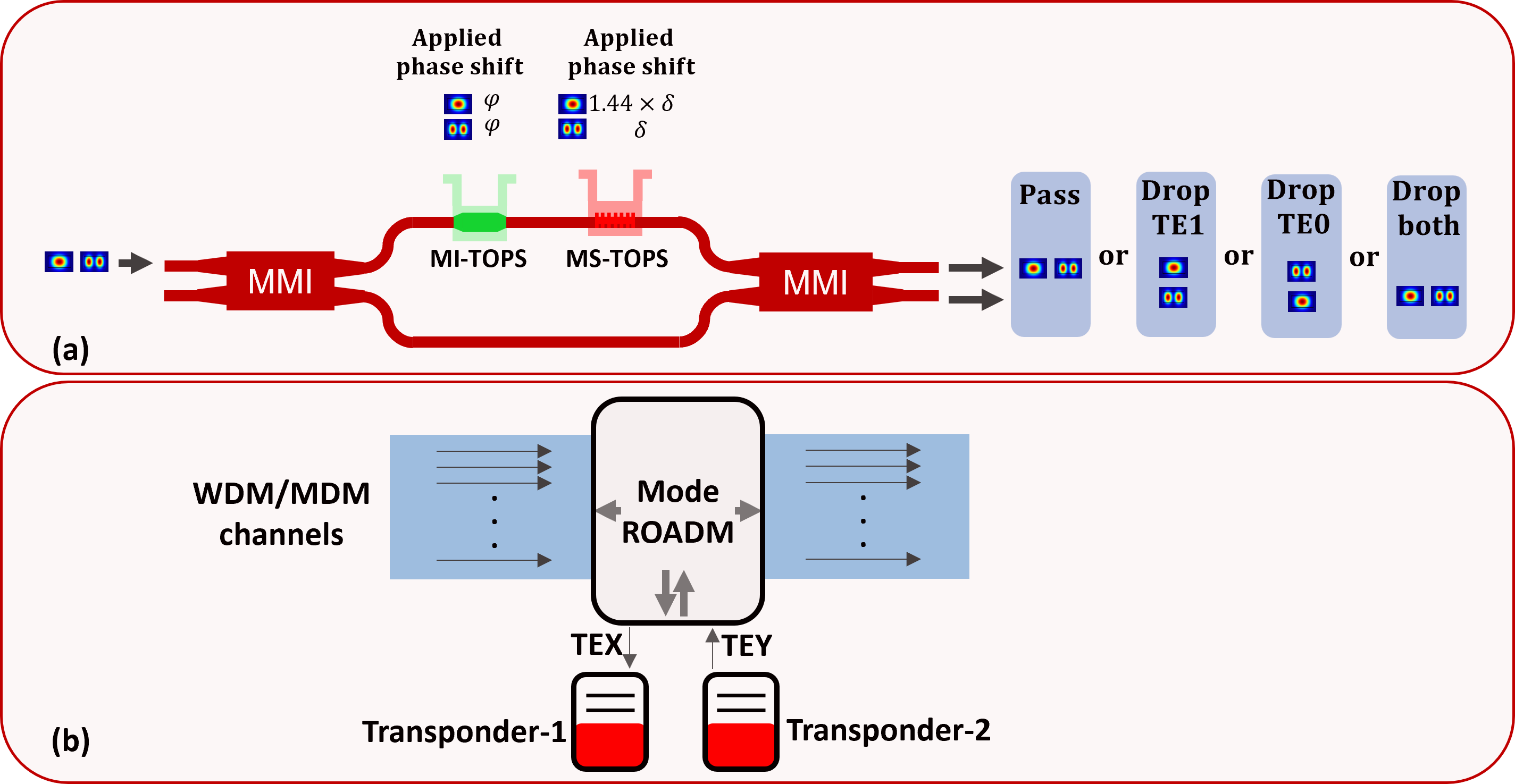}
\caption{(a) Schematic of a mode-selective switch used in a Mode-selective ROADM capable of adding/dropping TE0 or TE1. (b) Mode-selective ROADM in a WDM/MDM flexible system dropping the x\textsuperscript{th} TE mode channel and the y\textsuperscript{th} mode.}
\label{fig_mode_selective_switch_Schem}
\end{figure}

We recently introduced a mode-selective switch designed to route various modes to distinct ports without redistributing the modal fields. By cascading a MS-TOPS and a MI-TOPS within a MZI, we successfully demonstrate the selective addition and removal of TE0 or TE1 from a multimode bus waveguide in SiPh. Figure \ref{fig_mode_selective_switch_Schem}~(a) illustrates the configuration of this mode-selective switch. This switch takes the form of an MZI comprising two multimode interferometers serving as 3-dB splitter/combiners. Within one arm, two cascaded TOPS are employed. MI-TOPS imparts an identical phase shift to both TE0 and TE1, as detailed in section III. For the SWG duty cycle ratio of 0.4, the MS-TOPS applies a phase shift to TE0 that is approximately 1.44 times larger than that applied to TE1 \cite{JLT_Mohammad}. The phase shift applied to TE1 in the upper arm of MZI is the sum of the phase shifts applied by the two cascaded TOPS i.e., $\phi+\delta$. For TE0, the phase shift applied by the MS-TOPS is 1.44 times larger than that of TE1 leading to a total phase shift of $\phi+1.44\times\delta$.

By applying distinct biases to MI-TOPS and MS-TOPS, leading to different values of $\phi$ and $\delta$, we can achieve arbitrary and distinct phase shift values for TE0 and TE1. As depicted in fig. \ref{fig_mode_selective_switch_Schem}~(a), the mode-selective switch can be configured to pass or drop both modes, or selectively drop one of them. In a reversed configuration, it can also combine two modes applied to the two inputs of the MZIs. The SWG region in the MS-TOPS is designed to preserve the modal field distribution of the modes without converting them into supermodes, all while introducing unique temperature coefficients for TE0 and TE1. The design is fabricated at the Applied Nanotool (ANT) foundry using electron beam lithography, with a silicon layer thickness of 220 nm. The TOPSs utilize a 200 nm thick Titanium Tungsten (TiW) metal heater deposited on top of the 2.2 µm oxide cladding.

Fig. \ref{fig_mode_selective_switch_Schem}~(b) shows a potential application of the mode-selective switch in a Reconfigurable Optical Add/Drop Multiplexer (ROADM) system. ROADMs, introduced in the 2000s for WDM networks, responded to evolving network traffic demands and provided network flexibility \cite{ROADM_WDM}. They enable remote path modification of optical wavelength channels through wavelength-selective switches, allowing the addition and removal of specific wavelengths at a location in response to changes in traffic patterns. As a successor to WDM, the WDM/MDM network would benefit from a switch capable of adding/dropping individual modes. A mode-selective switch can perform this function in a WDM/MDM system.

\begin{figure}[!t]
\centering
\includegraphics[width=8.5cm]{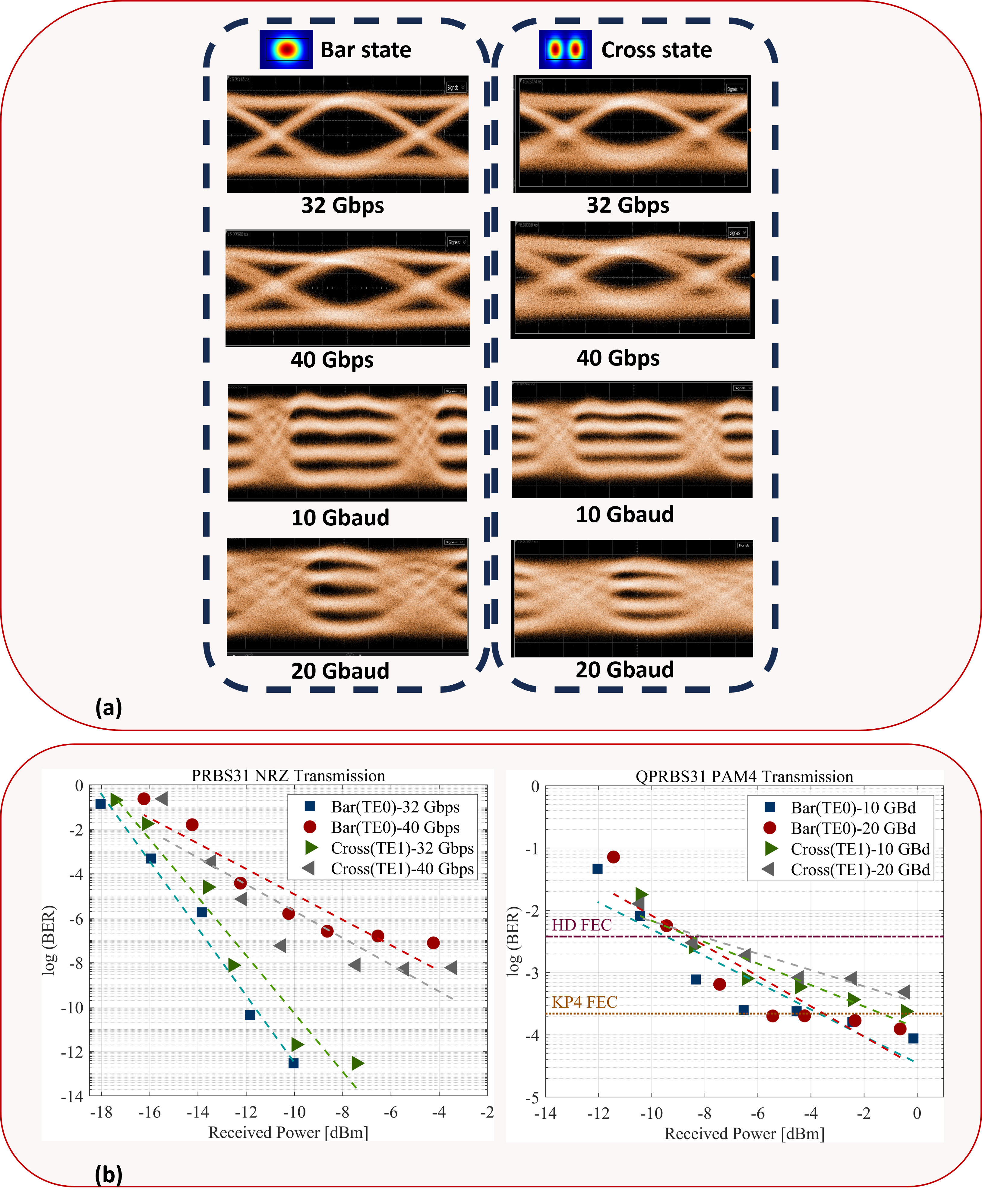}
\caption{(a) NRZ and PAM4 BER eye diagrams with a 100 mV/div y-axis for TE0 (Bar state output port) and TE1 (crossbar state output port). (b) The corresponding BER curves.}
\label{ROAM_results}
\end{figure}

In Figure \ref{ROAM_results}(a), eye diagrams for TE0 and TE1 transmission are displayed using PRBS31 NRZ and PRBS31Q PAM-4 for the two modes when TE0 is in the bar state and TE1 is in the cross state (dropped). Payload transmission is carried out with both modes simultaneously transmitted through the switch, resulting in degradation due to modal crosstalk. Corresponding bit error rate (BER) curves are presented in Figure \ref{ROAM_results}(b), demonstrating error-free NRZ transmission at 16 Gbps and a BER below $10^{-6}$ for 40 Gbps. For PAM-4 transmission, the BER remains below the KP4 forward error correction (FEC) limit of $2.2\times10^{-4}$ for TE0 and even below the HD FEC limit of $3.8\times10^{-3}$ for TE1 at 20 GBaud.

This work serves as a proof of concept for a mode-selective switch but was constrained by the single-etched SOI fabrication, resulting in grating couplers with 8 dB of coupling loss. This limitation led to challenging fiber-to-fiber insertion loss, subsequently impacting the BER measurement due to amplified spontaneous emission (ASE) noise from the required erbium doped fiber amplifier (EDFA) optical gain in the testbed. Additionally, coupling in and out of the chip occurs at the fundamental mode, with adiabatic mode (de)multiplexers converting higher-order modes to fundamental modes. Lower modal crosstalk in the mode-selective switch can be achieved through multimode couplers and few-mode fibers at the interface, eliminating the effects of mode multiplexers/demultiplexers.

\section{MDM in photonic computation}

While MDM has a decade-long history in optical communication, its relatively recent but burgeoning application in optical computing is gaining significant attention. This section delves into three primary applications of MDM. Firstly, MTMOP leverage the distinct propagation constants of various transverse modes to monitor phase shifts, facilitating precise phase adjustments essential for analog optical processors. Secondly, the Multi-Transverse-Mode Multiply-Accumulate Unit (MTM-MAC) capitalizes on the capability of adding different spatial modes through "non-coherent" summation. This approach addresses challenges in the coherent summation of conventional MACs, leading to a higher number of resolved output power. Lastly, we showcase the utilization of different transverse electric modes for encoding quantum information in Transverse-Mode Encoded Quantum Processors.

\subsection{Multi-Transverse-Mode MZI-based Optical Processors}

Programmable optical processors are promising structures for fast and energy-efficient analog optical computing. These processors can efficiently perform vector-matrix multiplication \cite{Farhad_IPJ,Farhad_JLT}, which currently consumes the largest portion of computation resources in machine learning and AI \cite{Compute_resources}. One main challenge related to the programmable optical processor is that, as analog devices, computation accuracy highly depends on programming accuracy and the phase setting of phase shifters. On the other hand, emerging applications for programmable optical processors ask for faster programming time so that the processor can be reconfigured to process multiple tasks. Calibration and programming of optical processors require sensing optical power and optical phase. Although sensing the optical power is easily feasible in photonic integrated circuits using on-chip photodetectors, sensing the optical phase requires more complex and elaborate hardware. One possible method for sensing/monitoring the phase shift applied by a phase shifter is to take advantage of different transverse electric modes as shown in MTMOPs \cite{Kaveh_JSTQE}. As discussed in section III, a MS-TOPS applies different phase shift to different modes but the ratio of phase shift values remains constant and it depends on the geometries of the MS-TOPS. For example for an SWG-based MS-TOPS shown in \cite{JLT_Mohammad}, for the SWG duty ratio of 40\%, the thermo-optic coefficient of TE0 is 44\% larger than that of TE1. We use this difference in thermo-optic coefficient in MTMOP to monitor the phase shift applied by the phase shifters. 

\begin{figure}[!t]
\centering
\includegraphics[width=8.5cm]{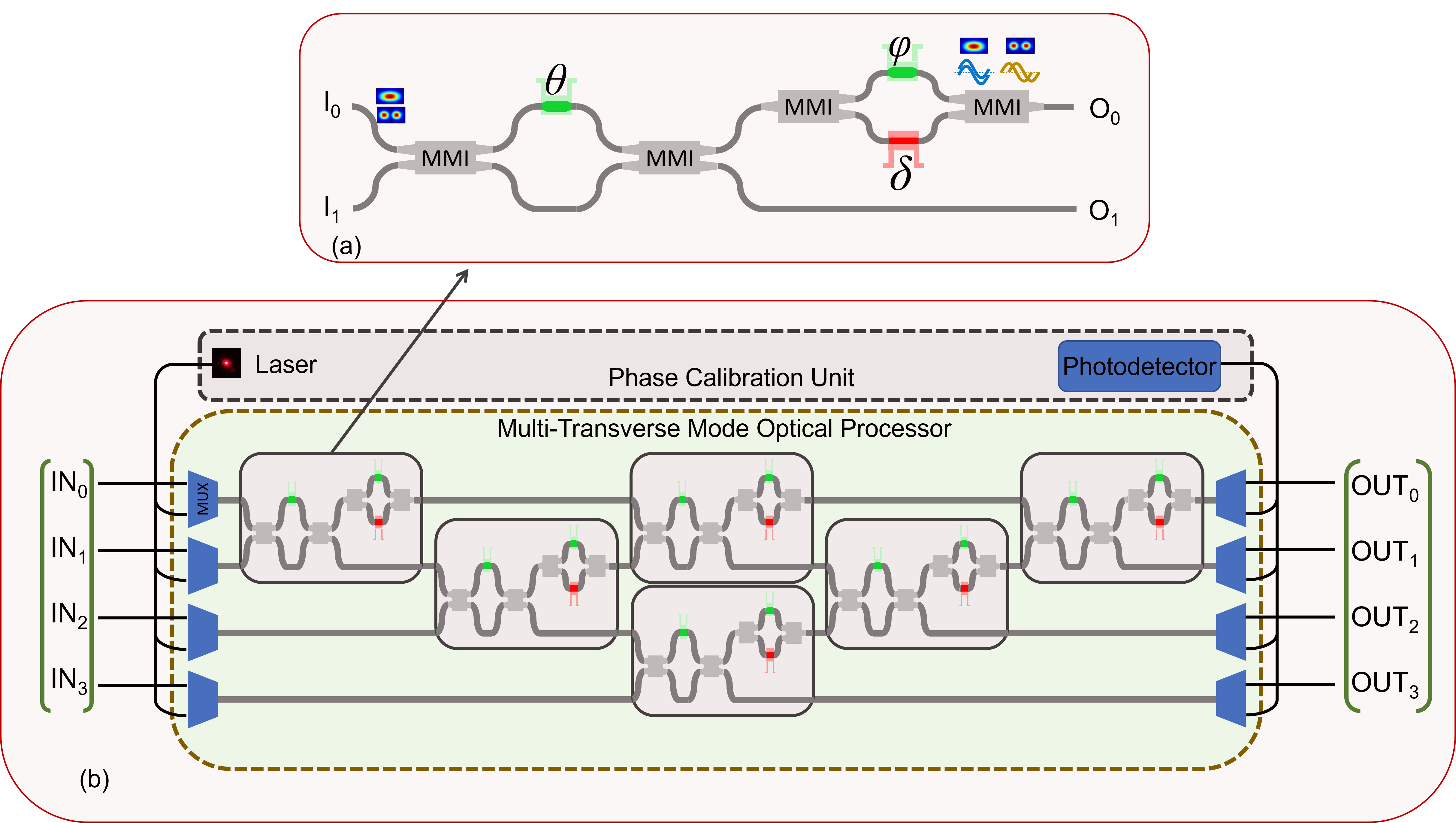}
\caption{(a) Schematic view of a 2 × 2 building block of MTMOP. TE0 is used as the main mode for carrying the data. TE1 is for monitoring the phase shifts in the calibration and programming phase. (b) 4 × 4 MTMOP based on the Reck architecture along with the phase calibration unit.}
\label{MTMOP}
\end{figure}

Figure~\ref{MTMOP}~(a) presents the proposed 2 × 2 building block schematic of the MTMOP. It uses two orthogonal TE optical modes: TE0 for carrying the main optical signal and TE1 for performing phase calibration. The internal phase shifter $\theta$ is a conventional MI-TOPS applying the same phase shift to the TE0 and TE1 modes. In the MTMOP, we replace the external phase shifter by an MZI composed of I) an MMI as a 50:50 beam splitter, II) two phase shifters ($\phi$ and $\delta$), III) a second MMI as the beam combiner. The phase shifter $\delta$ is MS-TOPS with different thermo-optic coefficient ($dn_{eff}/dT$) for TE0 and TE1. The phase shifter $\phi$ is MI-TOPS. The principle of operation is: we start with setting a bias to $\phi$ as an initial point for the desired TE0 phase shift. We then sweep the phase shifter $\delta$ bias voltage until the TE0 signal power at $O_0$ is maximized meaning the TE0 signal passing through the MZI including phase shifters $\phi$ and $\delta$ constructively interferes. This would not be the case for TE1 owing to the fact that the phase shift implied by $\delta$ is different for TE0 and TE1. Knowing $dn_{eff}/dT$ for TE0 and TE1, we calculate the phase shift applied to TE0 by measuring the transmission of TE1. In this way we can monitor the phase shift applied to TE0 by measuring the TE1 transmission. Knowing the phase shift applied to the TE0, we iterate the process until achieving the desired phase shift to TE0. Using this process, we can monitor the phase shift applied by the external phase shifters. Monitoring the phase shift is helpful in both calibration and programming phase of optical processors. We can calibrate the MTMOP completely on-chip using photodetectors and without need for coherent detection.  

Figure~\ref{MTMOP}~(b) shows a 4 × 4 MTMOP in the Reck architecture \cite{Reck} based on the aforementioned 2 × 2 building block. The idea presented in fig.~\ref{MTMOP}~(b) can be extended to other programmable optical processor architectures such as Clement mesh \cite{clements}, diamond mesh \cite{Diamond}, or Bokun mesh \cite{Bokun}. Details of programming techniques as well as discussion on scalability are presented in \cite{Kaveh_JSTQE}.

\subsection{Multi-Transverse-Mode Multiply-Accumulate Unit}

The terminology of the multiply-accumulate operation refers to the calculation of the product of two numbers and the addition of the result to an accumulator. 
A variety of linear mathematical operations, including matrix multiplications, convolutions, and dot products, could be represented by a series of MAC operations.
That explains why modern high-performance computing applications require a cornerstone of highly efficient MAC unit hardware in terms of energy and compute density.
Integrated photonics offers a promising solution for efficient MAC operations, demonstrating improvements over the electronic counterparts, toward realizing aj/MAC and PMACs/s/mm\textsuperscript{2} performance from both energy and compute density perspectives \cite{Bhavin_MAC}, \cite{Nikos_MAC}. 
A photonic MAC unit architecture based on microring weight banks drives microring resonators (MRRs) in and out of resonance to modulate the intensity of an input WDM signal, while a photodetector performs accumulation (weighted summation) through electro-optic conversion \cite{MRR_WB}.
System analysis suggests improved energy efficiency with an increase in the size of the MRR weight bank. However, network size scalability is predominantly limited by the number of WDM channels and crosstalk arising from the free spectral range (FSR) limitations \cite{Sudip_Sys_analysis}. 
While deploying two-point or contra-directional coupling schemes can suppress FSR constraints, they come at the expense of an additional heater, which deteriorates energy efficiency \cite{Two_P_Coupling_MRR}, \cite{FSR_Free_MRR}.
On the other hand, increasing the bit resolution not only demands more power-hungry drivers but is also limited by the finite amount of resonance extinction ratio.

Recent advancements in MDM components can potentially alleviate the above challenges by enabling signal modulation on fundamental and higher order transverse modes, while simultaneously deploying WDM technique \cite{MDM_2014_Lipson}. 
A mode-selective modulation design for interconnect applications utilized microring modulators to demultiplex modes from the multimode bus and multiplex the selected modes back to their originals \cite{Jia:19}. 
Such architecture for a multi-transverse-mode MAC unit application suffers from an increased insertion loss and modal crosstalk due to additional mode down and up conversions.
In \cite{Dai_Chip_Journal}, an MDM-based matrix-vector multiplier is proposed, where mode up and down conversion is implemented solely for efficient power splitting within the system. Weighting, on the other hand, is achieved exclusively in the fundamental mode, requiring coherent summation at the output.

\begin{figure}[!t]
\centering
\includegraphics[width=8.5cm]{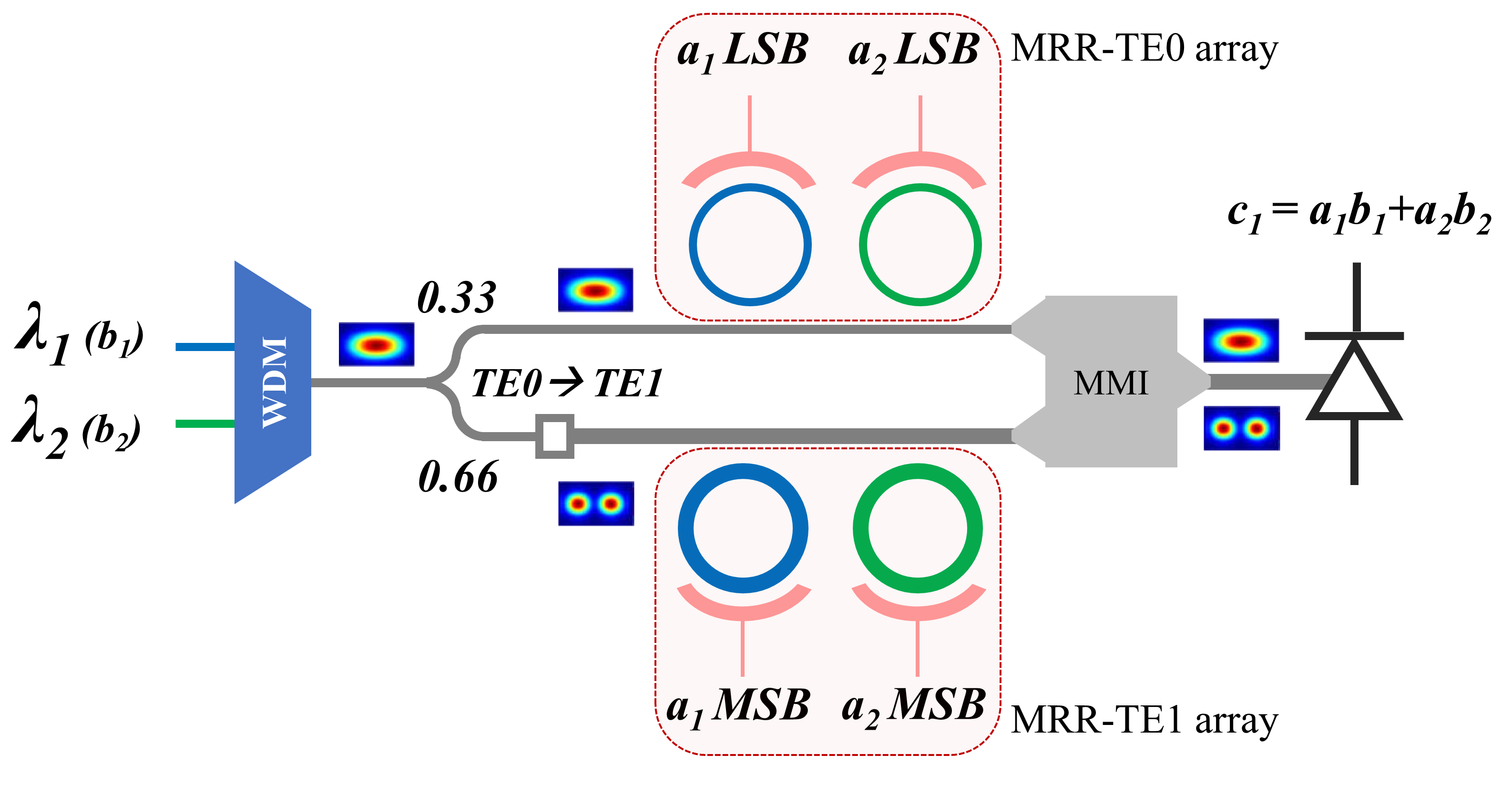}
\caption{A schematic representation of the proposed MTM-MAC unit for a two-channel weighted WDM input with $b_1$ and $b_2$ amplitudes. Fundamental (TE0) and first-order (TE1) transverse-mode MRRs are responsible for realizing the least (LSB) and the most (MSB) significant bits of each weight element ($a_1$ and $a_2$). The MDM building blocks are described in the text.}
\label{MTM_MAC}
\end{figure}

We recently introduced an MTM-MAC unit architecture that enables direct MRR-based weighting of the fundamental and first-order transverse-electric modes (TE0, TE1) as depicted in Figure \ref{MTM_MAC}. 
For simplicity, only two weighted WDM channels are considered with amplitudes of $b_1$ and $b_2$. 
The first-order mode (TE1) is excited on-chip using a mode converter.
Each element within the weight matrix (e.g., $a_1$ and $a_2$) is implemented using a pair of two MRRs coupled to the TE0 and TE1 signal paths, carrying an input optical power in a ratio of 0.33:0.66. 
The weighted TE0 and TE1 signals are directed to a 2×1 multimode interferometer. 
The orthogonality of the TE0 and TE1 optical signals allows for their non-coherent superposition in the MMI output path, without the need for any preceding phase-matching conditions.
The MMI output leads directly to the photodetector, which is inherently insensitive to modes. As a result, it accumulates both the weighted TE0 and TE1 signals, thereby completing an MTM-MAC operation.
Experimental validation of a single channel MTM-MAC unit in the SOI platform confirms that a binary weight control of each MRR-TE0 and MRR-TE1 resonating at the same wavelength produces a four-level resolved output power. 
A worst-case modal crosstalk of -12 dB is observed in the wavelength range of [1530:1550] nm.
This WDM-compatible design is scalable by further deployment of higher-order modes (TE2, TE3) leading to a more accurate MAC unit hardware.
The drawback of this design lies in the relatively large footprint of the deployed MRR-TE1 \cite{MRR_Winnie_2014}, which could be reduced by utilizing subwavelength-grating structures to minimize the coupling length \cite{MRR_TE1_SWG}.

\subsection{Transverse Mode Encoded Programmable Quantum Processor}

SiPh has been primarily utilized in integrated quantum computing to manipulate the quantum state of light and realize quantum gates \cite{Quantum_qianglarge}. Quantum information can be encoded in various degrees of freedom of single photons, including polarization, path, or wavelength. In recent years, path-encoded programmable quantum gates have been successfully demonstrated in SiPh, marking a significant step towards a large-scale integrated photonic quantum system \cite{Quantum_Jelmer}. However, encoding information in other degrees of freedom of single photons, such as polarization and transverse mode, is crucial for the scalability of the system. Recently, attempts have been made to encode information in transverse electric modes and manipulate it using transverse mode integrated quantum gates \cite{Quantum_Dai}. Although innovative and efficient, the design presented in  \cite{Quantum_Dai} is not programmable. Consequently, it can only perform a specific operation, in that case, Controlled Not (CNOT). Additionally, since the design is not programmable, fabrication variations cannot be compensated for through fine-tuning the phase shifters.

\begin{figure}[!t]
\centering
\includegraphics[width=8.5cm]{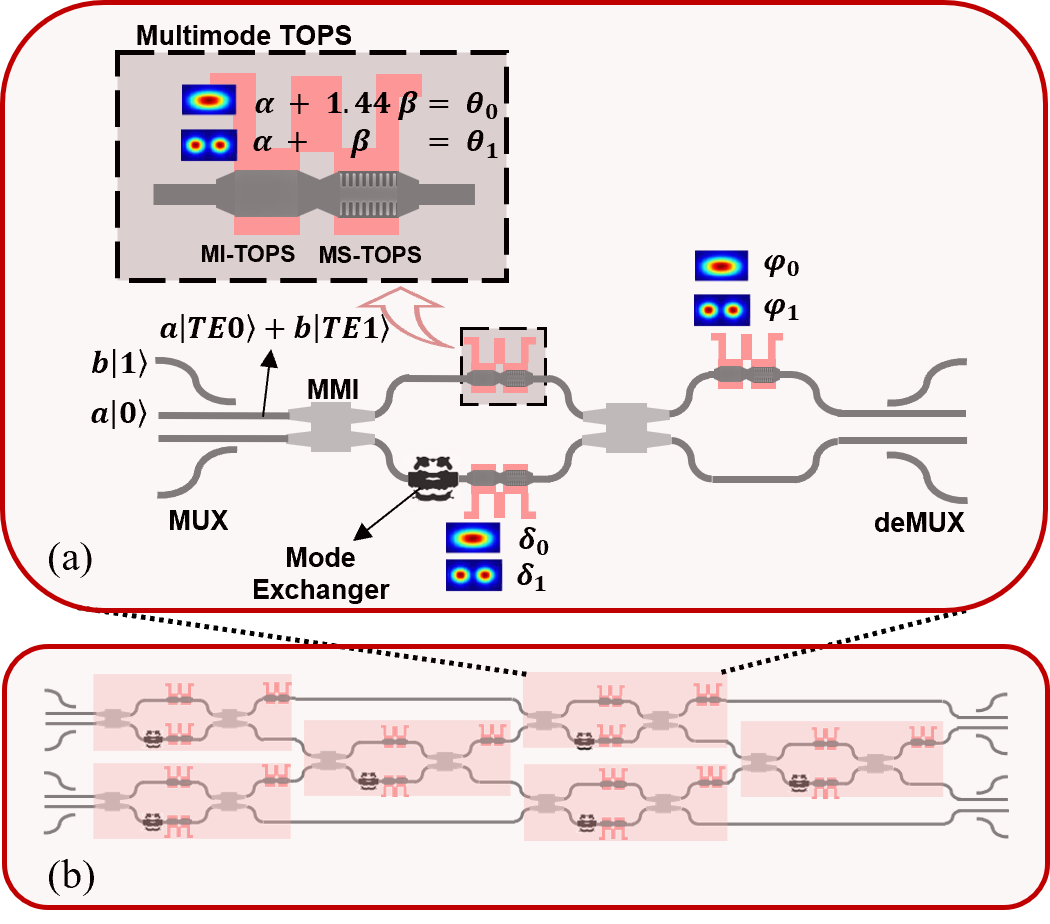}
\caption{
(a) A schematic illustration of a 4 × 4 building block for the Transverse Mode Encoded Programmable Quantum Processor, including three multimode TOPS. Each multimode TOPS is comprised of a MI-TOPS and MS-TOPS. (b) Scaling the processor to 8 × 8 using Clements architecture.
}
\label{MTM_Quantum}
\end{figure}

Figure~\ref{MTM_Quantum} depicts a schematic view of a 4 × 4 building block of a programmable transverse-mode encoded processor that we propose, leveraging TE0 and TE1 to encode quantum information. The primary advantage of the proposed transverse mode encoded programmable quantum processor is its ability to independently manipulate different transverse modes without converting them to the fundamental mode.
All the components used in this block, as shown in fig.~\ref{MTM_Quantum}~(a), are available in the \cite{PDK} library. The key structure in this block is an MZI with two multimode input waveguides, each carrying two modes, TE0 and TE1. In the input of the MUX, we have a generic path-encoded qubit represented as $a |0\rangle +b |1\rangle$, where $|0\rangle$ and $|1\rangle$ are two basis states corresponding to the presence of a single photon on the first and second single-mode waveguides, respectively.
The single-mode waveguides are 0.5~\textmu m wide, allowing only the propagation of the fundamental TE0 mode. The MUX converts the path-encoded state to a coherent superposition of TE0 and TE1 transverse electric mode states, $a |TE0\rangle +b |TE1\rangle$, of a single photon propagating in a multi-mode waveguide where $a |TE0\rangle$ and $b |TE1\rangle$ are the basis states corresponding to the propagation of the single photon on TE0 and TE1 modes, respectively.
The multimode waveguide is 0.96~\textmu m wide, allowing the propagation of TE0 and TE1. We use mode-insensitive MMIs as splitters and combiners of the MZI. The processor utilizes three multimode TOPSs: $\delta$, $\phi$, and $\theta$. Each multimode TOPS comprises cascaded MI-TOPS and MS-TOPS, providing two degrees of freedom to independently control the phase shifts applied to TE0 and TE1, similar to the approach we discussed for the mode ROADM in Fig.~\ref{fig_mode_selective_switch_Schem}. The lower arm includes a mode exchanger to swap TE0 and TE1, providing a path for transmission between TE0 to TE1 and vice versa.

Applying the method discussed in \cite{Farhad_book}, we define the transfer matrix of the structure by multiplying the corresponding transfer matrices of the cascaded blocks. For a 50:50 splitting ratio of combiners and splitters, the linear transformation matrix of the input optical field to the output optical field is:
\begin{eqnarray}
\frac{1}{2}\times
\begin{bmatrix}
e^{j\phi_0+j\theta_0} & -e^{j\phi_0+j\delta_0} & je^{j\phi_0+j\theta_0} & je^{j\phi_0+j\delta_0}\\
-e^{j\phi_1+\delta_1} & e^{j\phi_1+\theta_1} & je^{j\phi_1+\delta_1} & je^{j\phi_1+\delta_1}\\
je^{j\theta_0} & je^{j\delta_0} & -e^{j\theta_0} & e^{j\delta_0}\\
je^{j\delta_1} & je^{j\theta_1} & e^{j\delta_1} & e^{j\theta_1}\\
\end{bmatrix} 
\end{eqnarray}

Modifying the values of the three TOPS based on the transformation matrix in eq. (4), we can perform a unitary transformation on the optical fields of the two paths and two orthogonal modes. The presented design performs a 4 × 4 unitary transformation employing only a single MZI. We can use the building block presented in Fig.~\ref{MTM_Quantum}~(a) on a larger mesh as shown in Fig.~\ref{MTM_Quantum}~(b) to realize a linear transformation on larger dimensions. In the presented multi-transverse-mode processor, the number of MZIs required for an N × N transformation on Clements mesh is (N(N-2))/8, which is asymptotically one-fourth of the MZIs required in a single-mode path-encoded system (i.e., (N(N-1))/2) \cite{clements}. The mesh depth (number of consecutive MZIs in the longest path) is also half that of a single-mode path-encoded system.

\section{Conclusion}
In this work, we have presented a comprehensive library of MDM components, including MI-TOPS, MS-TOPS, MMIs, mode (de)Mux, mode converters, and mode exchangers. These components are developed using classical and inverse design and are all compatible with standard 220~nm SiPh foundries. We have also discussed how MDM can contribute to on-chip and chip-to-chip communication, providing increased throughput by exploiting higher-order modes. The mode ROADM presented lays the groundwork for flexible MDM architectures in the future. Additionally, we have explored MDM applications in computing, including MTMOP for faster and more accurate programming of optical processors. The discussion on MTM-MAC highlights its capability for non-coherent summation of multiple TE modes, offering scalability in the number of bits for photonic MACs. Furthermore, the transverse mode encoded programmable quantum processor has been introduced, providing an extra degree of freedom for encoding quantum information in SiPh.


%

\section*{Acknowledgment}

Authors acknowledge the help from Canadian Microelectronic Corporation (CMC) for the subsidized multiproject wafer fabrication through Applied Nanotools (ANT) as well as financial support from the Natural Sciences and Engineering Research Council of Canada (NSERC) and Fonds de recherche du Québec (FRQNT). 

\ifCLASSOPTIONcaptionsoff
  \newpage
\fi



\bibliographystyle{IEEEtran}
%

%

\begin{IEEEbiography}[{\includegraphics[width=1in,height=1.25in,clip,keepaspectratio]{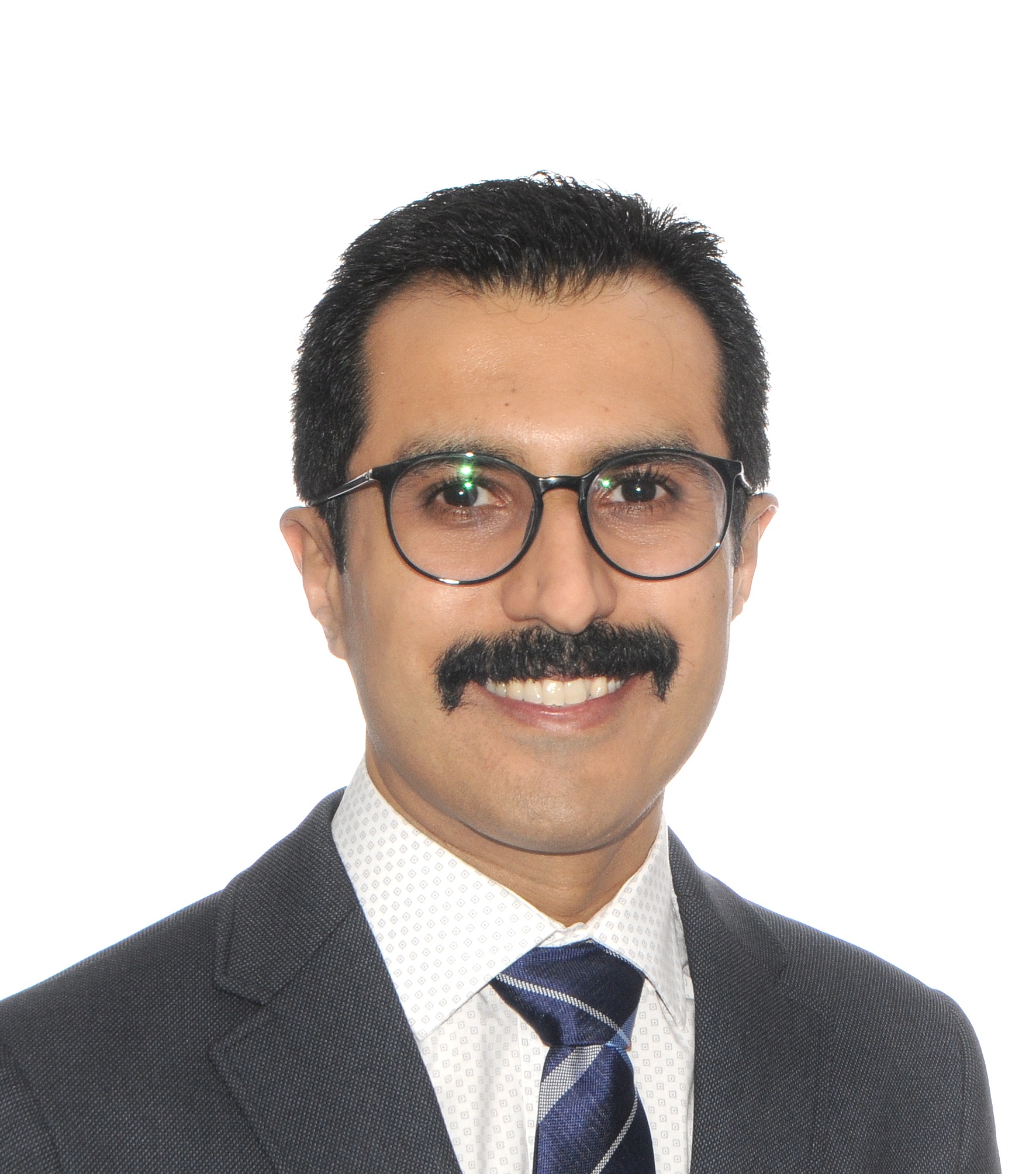}}]{Kaveh~(Hassan)~Rahbardar~Mojaver}
(M'17) received the B.S. and M.S. degrees from Amirkabir University of Technology (Tehran Polytechnic) in 2009 and 2011, respectively, and the Ph.D. degree from Concordia University in 2018, all in electrical engineering. From 2018 to 2023, he served as a postdoctoral researcher at the Photonic DataCom team at McGill University. In 2023, he joined Colorado State University as an assistant professor. His research focuses on photonic integration for optical computing, quantum computing, and data communications.
From 2014 to 2018, he was a part of the Reliable Electron Device Laboratory at Concordia University, where he conducted research on micro-fabrication and physics-based modeling of III-nitride heterojunction field-effect transistors. From 2018 to 2023, as a postdoctoral researcher, Dr. Mojaver conducted research on photonic integrated circuits, mainly focusing on silicon photonics for optical communications and computations.
Dr. Mojaver’s research works have been published in several journal and conference publications. He has received the Fonds de Recherche du Québec – Nature et technologies (FRQNT) postdoctoral scholarship, Concordia Merit scholarship, Concordia Accelerator Award, and STARaCom scholarship.
\end{IEEEbiography}

\begin{IEEEbiography}[{\includegraphics[width=1in,height=1.25in,clip,keepaspectratio]{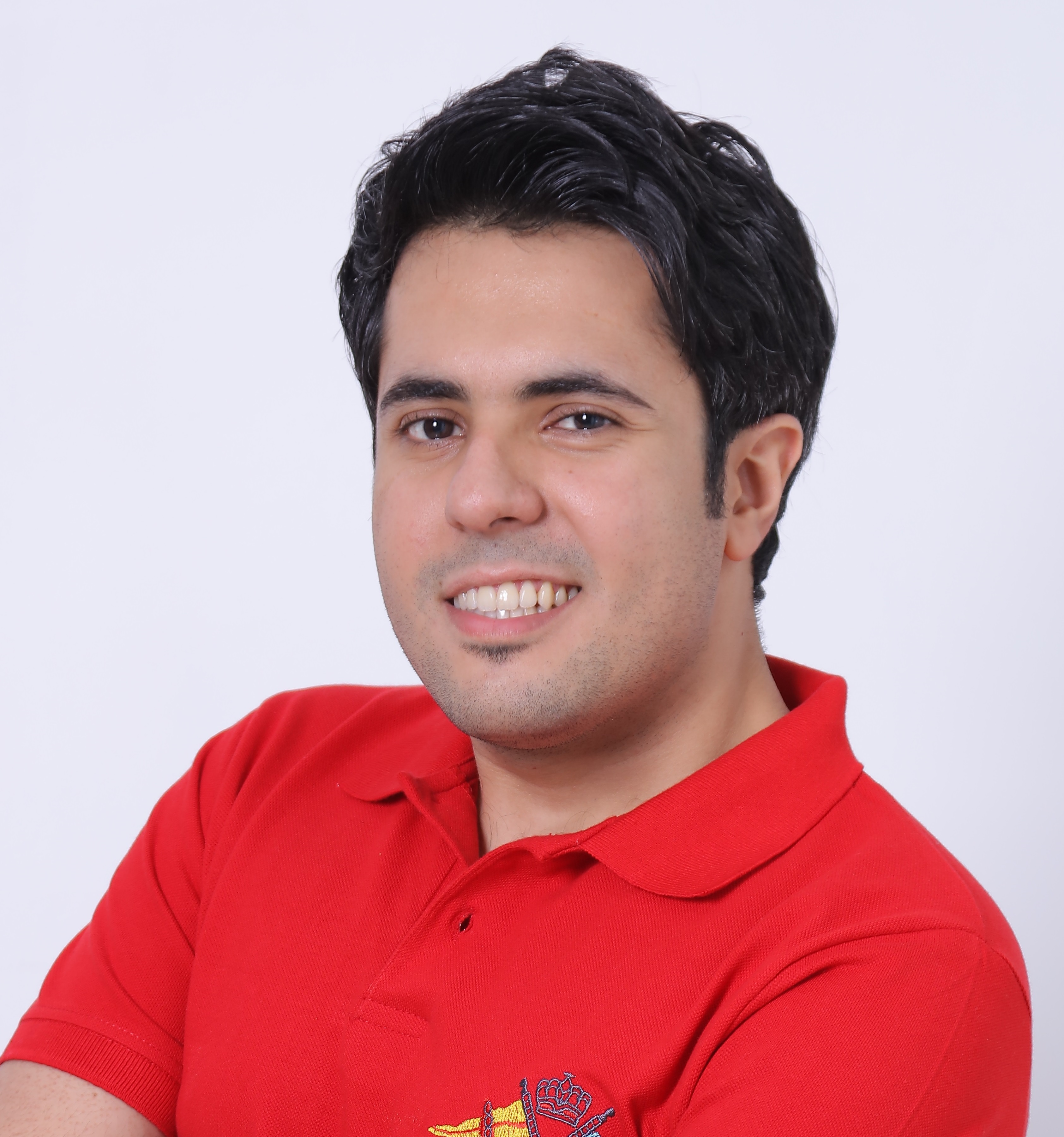}}]{S.~Mohammad~Reza~Safaee}
(M’22)
received both his B.S. and M.S. degrees
from the Department of Electrical
Engineering at Iran University of Science
and Technology, Tehran, Iran, in 2013 and
2015, respectively, with a major in
electronics. He is currently a Ph.D.
candidate at McGill University, affiliated with the Photonics
Systems Group within the Department of Electrical and
Computer Engineering, located in Montreal, Quebec, Canada.
His ongoing research primarily focuses on developing
building blocks that enable on-chip optical computing applications, programmable photonics, and photonic digital to analog converters.
\end{IEEEbiography}

\begin{IEEEbiography}[{\includegraphics[width=1in,height=1.25in,clip,keepaspectratio]{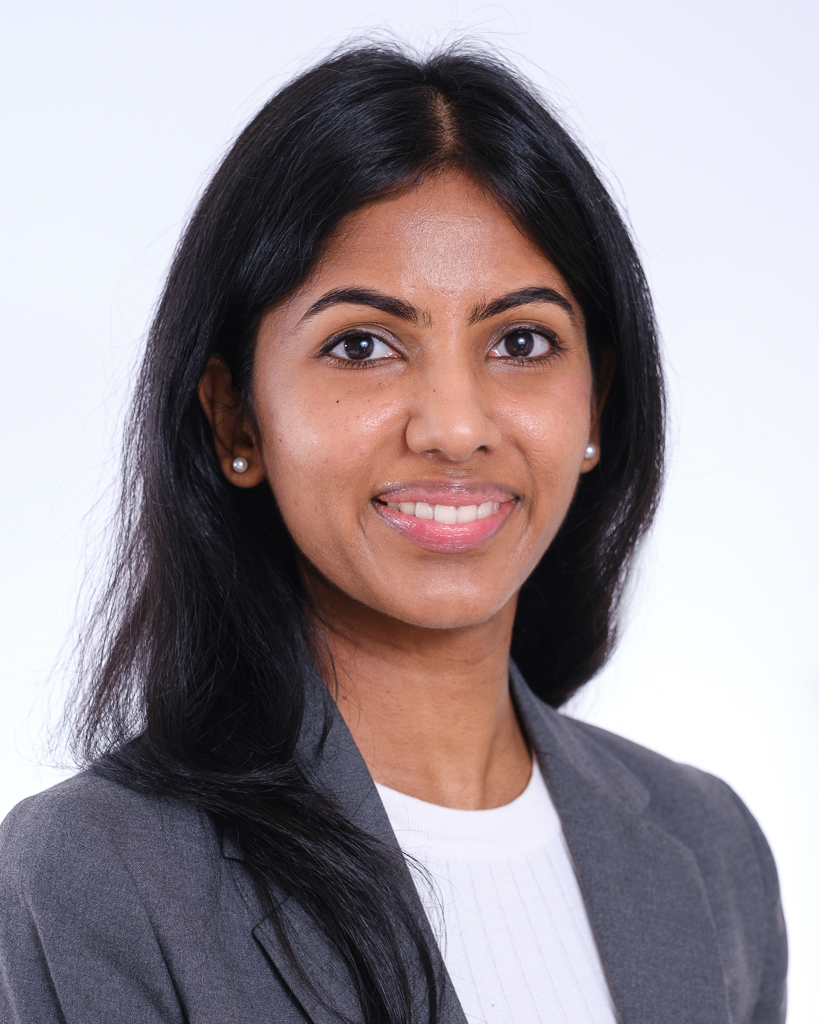}}]{Sunami~Sajjanam~Morrison}
(M'19)
 received her B.Eng in Electronics and Communication Engineering at Anna University, Chennai, India, in 2014, she then went on to obtain her M.S in Electrical and Computer Engineering with Engineering Practice from Boston University in 2016 focusing on photonics. She is currently a Ph.D. Candidate in the Electrical and Computer Engineering Department at McGill University. She is a part of the Photonics Systems Group, and her speciality lies in the development and optimisation of high-speed silicon photonic based transmitters for short-reach communication.
\end{IEEEbiography}

\begin{IEEEbiography}[{\includegraphics[width=1in,height=1.25in,clip,keepaspectratio]{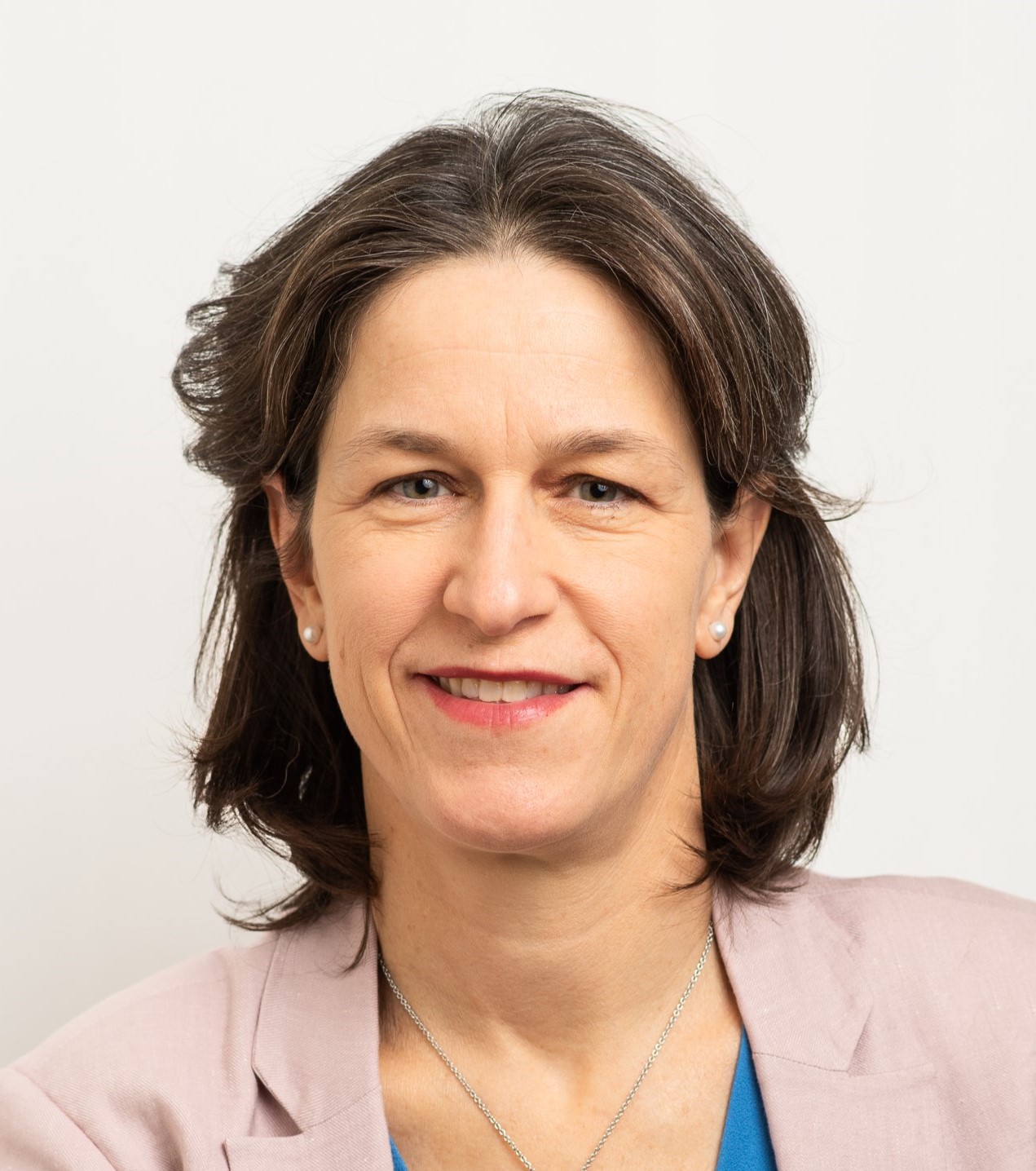}}]{Odile~Liboiron-Ladouceur}
(M’95, SM’14) received the B.Eng. degree in
electrical engineering from McGill
University, Montreal, QC, Canada, in
1999, and the M.S. and Ph.D. degrees in
electrical engineering from Columbia
University, New York, NY, USA, in 2003
and 2007, respectively. She is currently an
Associate Professor with the Department of Electrical and
Computer Engineering, McGill University. She was the graduate program director of the department (2017-2023). She was an
associate editor of the IEEE Photonics Letter (2009–2016) and
was on the IEEE Photonics Society Board of Governance
(2016–2018). She holds seven granted U.S. patents, and two provisional patents, and coauthored
over 100 peer-reviewed journal papers and 150 papers in
conference proceedings. She gave over 30 invited talks on the
topic of photonics for computing, photonic computational
design methodologies, and photonic-electronic co-design. She
is the 2018 recipient of McGill Principal’s Prize for
Outstanding Emerging Researcher and the 2023 William and
Rhea Seath award for engineering innovation.
\end{IEEEbiography}


\vfill


\end{document}